# Anomalous thermodiffusion, absolute negative mobility and reverse heat transport in a single quantum dot

Yanchao Zhang[*], Xiaolong Lü

*School of Science, Guangxi University of Science and Technology, Liuzhou 545006, People's Republic of China*

We investigate the steady-state transport characteristics of a quantum dot system consisting of a single energy level embedded between two reservoirs under the influence of both the temperature gradient and bias voltage. Within tailored parameter regimes, the system can exhibit three counterintuitive transport phenomena of anomalous thermodiffusion, absolute negative mobility and reverse heat transport respectively. These counterintuitive phenomena do not violate the second law of thermodynamics. Moreover, absolute negative mobility and reverse heat transport can be identified by a reversible energy level. These anomalous transports are different from thermoelectric transports and provide different perspectives for a more comprehensive understanding of the transport characteristics of quantum systems.

[*] Email: zhangyanchao@gxust.edu.cn

## I. INTRODUCTION

In equilibrium thermodynamics, temperature gradient may induce particle transport, even in the absence of concentration gradient. Normally, the direction of the particle transport is the same as the direction of the temperature gradient, i.e., from high-temperature heat reservoir to low-temperature heat reservoir, this phenomenon is called thermodiffusion, also known as thermophoresis or Soret effect [1-4]. However, a somewhat counterintuitive phenomenon of particle flows moving against a temperature gradient known as anomalous thermodiffusion (AT) has also been found in different systems in recent years [5-7]. This counterintuitive transport phenomenon has stimulated new and extensive interest in exploring the novel transport behavior of thermodynamic systems.

When a system is in thermal equilibrium, the response of the system to an applied force is always in the same direction as the applied force, in order to towards a new equilibrium. Therefore, the it is impossible for the system's response to be opposite to an applied force, which would violate the laws of thermodynamics. When the system deviates from thermal equilibrium under the action of the temperature gradient, the response of the system becomes possible in the opposite direction of the applied force. A typical example is the thermoelectric effect, where an applied force, i.e., the bias voltage, is applied to the system and acts in the opposite direction to the temperature gradient [8]. Based on the Seebeck effect, the particle current can move against the bias voltage driven by the temperature gradient, or based on the Paltier effect, the heat flow can flow from low-temperature heat reservoir to high-temperature heat reservoir against the temperature gradient driven by the bias voltage. The thermodynamic properties of these processes have been extensively studied in thermoelectric heat engines [9, 10] and thermoelectric refrigerators [11, 12].

However, when the applied force is in the same direction as the temperature gradient, an inverse response implies an induced current (e.g., either particle or heat) against both applied forces, which is highly counterintuitive, although there is no fundamental law that forbids abnormal phenomenon. In the study of particle transport, a counterintuitive behavior called absolute negative mobility (ANM), in that the steady state particle current of the system is opposite to the thermodynamic force. In recent years, the ANM has been investigated in a variety of nonequilibrium systems, such as Brownian particles [13-20], semiconductor superlattices [21], microfluidic systems [22], Josephson junctions [23], one-dimensional interacting Hamiltonian system [24], Coulomb-coupled quantum dots system [25], active polymer [26], and active tracer particle [27, 28]. Another counterintuitive behavior related to the heat transport is known as reverse heat transport (RHT), which heat can flow from a low-temperature heat reservoir to a high-temperature heat reservoir against all thermodynamic forces. Recently, the RHT have also been found in both quantum [25, 29, 30] and classical systems [24].

In this paper, we focus on the steady-state transport characteristics of a single-level quantum dot system deviates from the equilibrium when driven by temperature gradient and bias voltage. It can be found that in tailored parameter regimes, the system will exhibit counterintuitive transport phenomena, which are anomalous thermodiffusion, absolute negative mobility and reverse heat transport, respectively. This paper is organized as follows. In Sec. II, the model and theory are detailed. In Sec. III, the AT phenomena is described. In Sec. IV, the ANM induced by thermal equilibrium fluctuation is analyzed. In Sec. V, the RHT is presented and the thermodynamic characteristics of the anomalous transport are analyzed. In Sec. VI, the thermoelectric transport is briefly discussed. In Sec. VII, the conclusions are summarized.

## II. MODEL AND THEORY

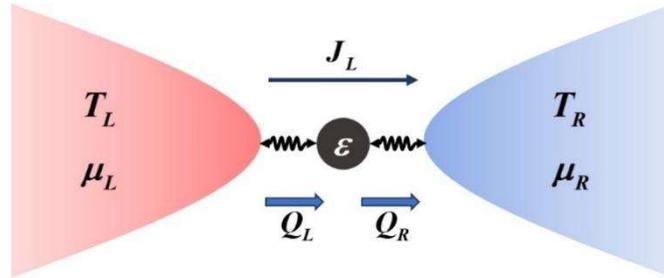

FIG. 1 The schematic diagram of a single quantum dot embedded between two reservoirs at different temperatures and chemical potentials.

We consider a simplest prototype model that consists of a quantum dot with a single resonant energy level $\varepsilon$ in contact with two reservoirs at different temperatures $T_\nu$ and chemical potentials $\mu_\nu$ $(\nu = L, R)$, as shown schematically in Fig. 1. In the sequential tunneling regime, the quantum dot is either empty or filled, described as occupation probabilities $p_0$ and $p_1$. Thus, the time evolution of occupation probabilities is given by a master equation

$$\frac{d}{dt}\begin{pmatrix} p_0 \\ p_1 \end{pmatrix} = \begin{pmatrix} -W_{10} & W_{01} \\ W_{10} & -W_{01} \end{pmatrix}\begin{pmatrix} p_0 \\ p_1 \end{pmatrix}, \tag{1}$$

and the transition rates are given by Fermi's golden rule

$$W_{10} = \sum_{\nu=L,R} W_{10}^\nu = \sum_{\nu=L,R} \gamma_\nu f_\nu, \tag{2}$$

$$W_{01} = \sum_{\nu=L,R} W_{01}^\nu = \sum_{\nu=L,R} \gamma_\nu (1-f_\nu), \tag{3}$$

where $f_\nu = [1+e^{(\varepsilon-\mu_\nu)/k_B T_\nu}]^{-1}$ is the Fermi-Dirac distribution function and $\gamma_\nu$ is bare tunneling rate. In the steady state, i.e., $dp_i/dt = 0$ $(i=0,1)$, the occupation probabilities can be solved as

$$p_0 = \frac{W_{01}}{W_{01}+W_{10}}, \tag{4}$$

$$p_1 = \frac{W_{10}}{W_{01}+W_{10}}. \tag{5}$$

The steady-state current from the left reservoir to the quantum dot is then given by

$$J_L = W_{10}^L p_0 - W_{01}^R p_1 = \gamma(f_L - f_R), \tag{6}$$

where $\gamma = \gamma_L \gamma_R / (\gamma_L + \gamma_R)$, and $J_R = W_{01}^R p_1 - W_{10}^L p_0 = J_L$.

The steady-state heat flow from left reservoir and the steady-state heat flow into right reservoir are respectively given by

$$Q_L = (\varepsilon - \mu_L) J_L, \tag{7}$$

$$Q_R = (\varepsilon - \mu_R) J_R, \tag{8}$$

According to stochastic thermodynamics, the entropy production rate associated with the master equation is given by

$$s = \sum_{i,j,\nu} W_{ij}^{\nu} p_j \ln \frac{W_{ij}^{\nu} p_j}{W_{ji}^{\nu} p_i}, \qquad (9)$$

This is consistent with the standard thermodynamic statement, i.e., $s = Q_R/T_R - Q_L/T_L$.

In our next numerical simulations, we set $T_L = T + \Delta T/2$ and $T_R = T - \Delta T/2$, where $T = (T_L + T_R)/2$ is the average temperature of the system, and $\Delta T = T_L - T_R$ is the temperature gradient. When $\Delta T > 0$, i.e., the left reservoir is at higher temperature, which is the forward configuration with the temperature gradient. $\mu_L = \mu + e\Delta V/2$ and $\mu_R = \mu - e\Delta V/2$, where $\Delta V = (\mu_L - \mu_R)/e$ is bias voltage, and $\Delta V > 0$ indicates the forward bias voltage. We set $\mu = 0$ as reference and other parameters use natural units, i.e., $k_B = e = \gamma = 1$.

### III. ANOMALOUS THERMODIFFUSION

We first consider a simple case with $\Delta T > 0$ and $\Delta V = 0$, that is, the system only applies one positive temperature gradient. The curves of the steady-state current $J_L$ and the steady-state heat flow $Q_L$ with a single energy level $\varepsilon$ at different system temperatures $T$ are shown in Fig. 2. It can be found that the steady-state heat flow $Q_L$ is always greater than zero, that is, the heat flows from the high-temperature to the low-temperature heat reservoirs, which is the expected result. For the steady-state current $J_L$, when $\varepsilon > 0$, $J_L > 0$, i.e., the particles will flow from the high-temperature to the low-temperature heat reservoirs. However, when $\varepsilon < 0$, the current of the particles will reverse and flow from the low-temperature to the high-temperature heat reservoirs. It is clearly that thermodiffusion of particles in single quantum dot system exhibits a counterintuitive phenomenon, where the particles move against temperature gradient $\Delta T$ and accumulate in high-temperature heat reservoir. This phenomenon is called the anomalous thermodiffusion. In Fig. 2, it can also be seen that as long as the single energy level $\varepsilon < 0$, the AT phenomena will appear regardless of the average temperature of the system.

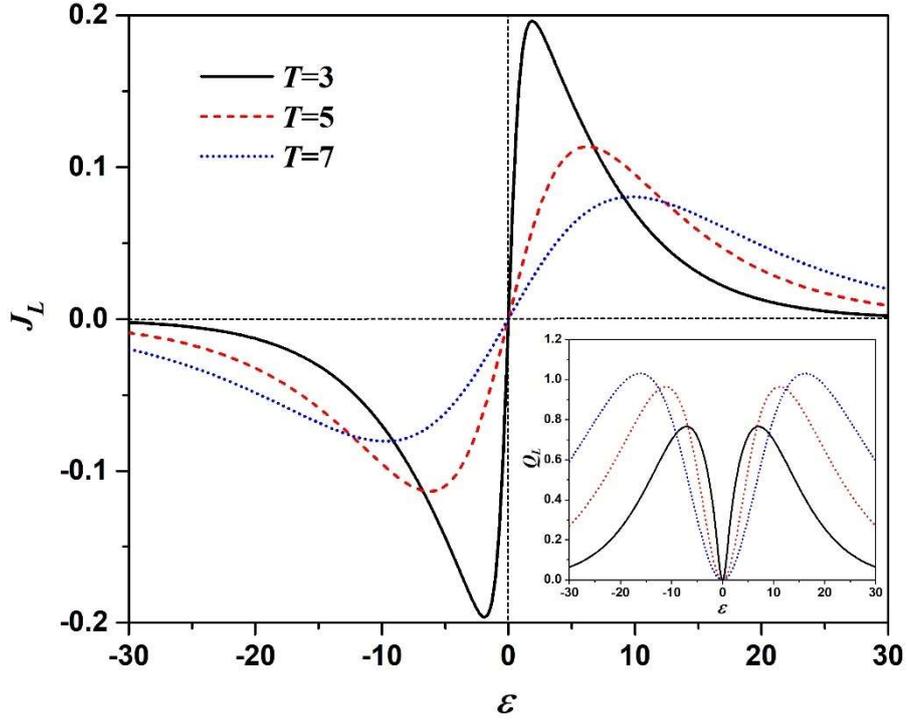

FIG. 2 The steady-state current $J_L$ is depicted as a function of a single energy level $\varepsilon$ at different system temperatures $T$. The anomalous thermodiffusion occurs when $\varepsilon < 0$. Inset: The steady-state heat flow $Q_L$ as a function of a single energy level $\varepsilon$ at different system temperatures $T$. The parameters $\Delta T = 5$, $\Delta V = 0$.

The steady-state current $J_L$ and the steady-state current $Q_L$ are depicted as a function of a single energy level $\varepsilon$ at different temperature gradients $\Delta T$, as shown in Fig. 3. It is found that as the temperature gradient $\Delta T$ increases, both the steady-state current $J_L$ and the steady-state heat flow $Q_L$ of the system are enhanced. At the same time, the AT phenomenon of the system will become more significant.

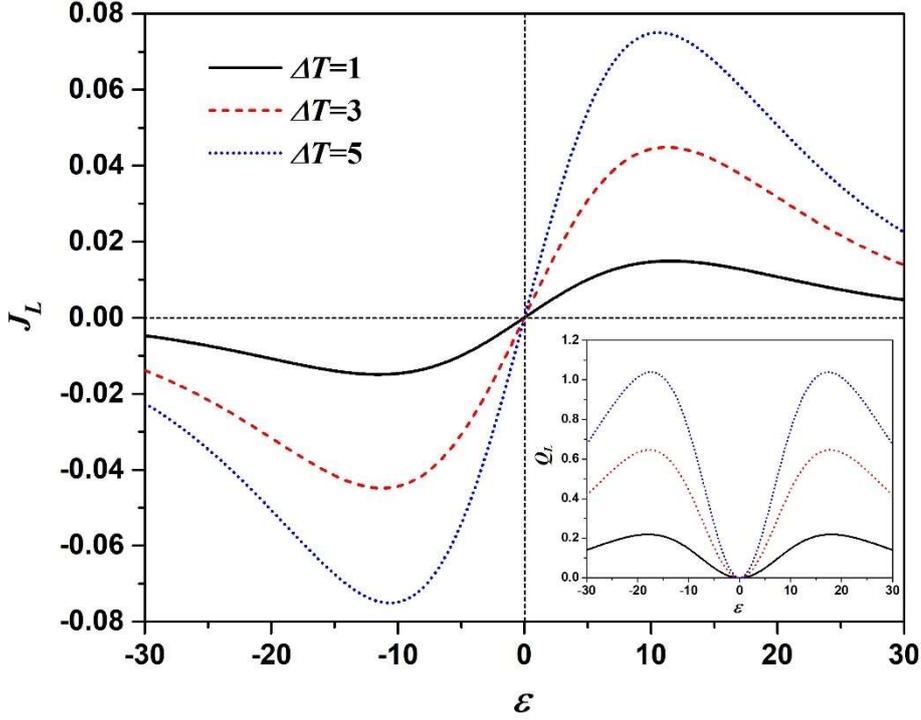

FIG. 3 The steady-state current $J_L$ is depicted as a function of a single energy level $\varepsilon$ at different temperature gradients $\Delta T$. The anomalous thermodiffusion occurs when $\varepsilon < 0$. Inset: The steady-state current $Q_L$ as a function of a single energy level $\varepsilon$ at different temperature gradients $\Delta T$. The parameters $T = 7.5$, $\Delta V = 0$.

## IV. ABSOLUTE NEGATIVE MOBILITY

We now consider the transport characteristics of the steady-state current $J_L$ in the presence of a forward bias voltage, i.e., $\Delta V > 0$. The steady-state current $J_L$ is depicted in Fig.4 as a function of a single energy level $\varepsilon$ at different temperature gradients $\Delta T$. It can be found that when the temperature gradient $\Delta T = 0$, that is, the system is in thermal equilibrium, the steady-state current $J_L$ is always greater than zero, that is, it flows in the direction of bias voltage. When the system deviates from the thermal equilibrium state and a positive temperature gradient $\Delta T > 0$ is superimposed along the direction of bias voltage, it is incredible that the steady-state current $J_L$ of the system will be less than zero, that is, the steady-state current $J_L$ flows backward against a bias voltage. This phenomenon is called absolute negative mobility. Obviously, the ANM is induced by thermal equilibrium fluctuation, and essentially, the ANM can be attributed to the AT that can be generated by temperature gradient overcoming bias voltage when the bias voltage is not large.

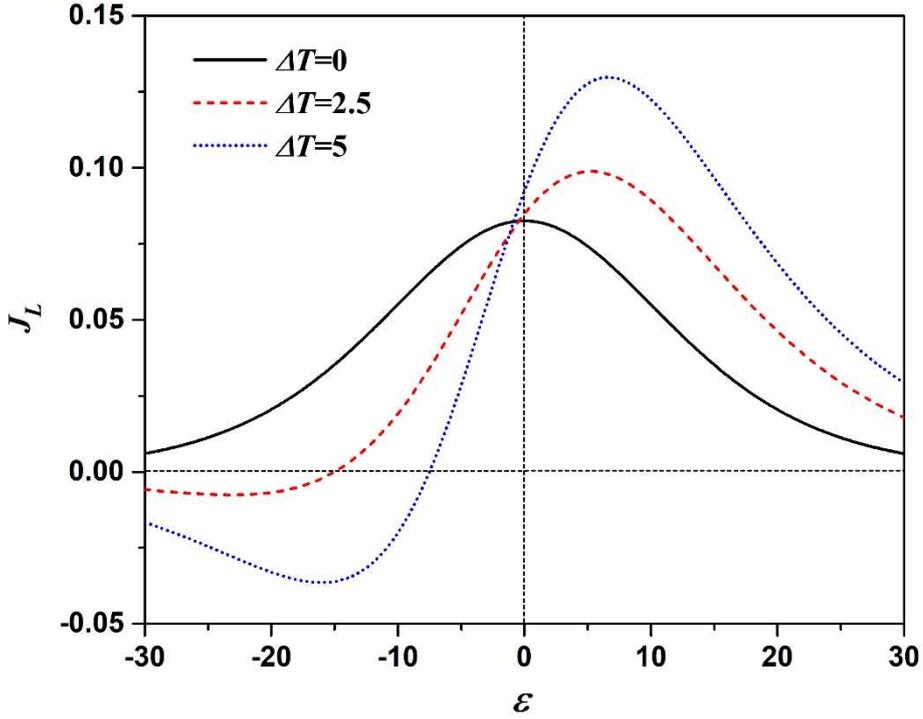

FIG. 4 The steady-state current $J_L$ is depicted as a function of a single energy level $\varepsilon$ at different temperature gradients $\Delta T$. The absolute negative mobility occurs in regions where $\varepsilon < 0$ when $\Delta T > 0$. The parameters $T = 7.5$, $\Delta V = 5$.

Fig. 5 shows the steady-state current $J_L$ as a function of the temperature gradient $\Delta T$ (a) and the bias voltage $\Delta V$ (b), respectively. It can be found in Fig. 5 (a) that in the range of $0 < \Delta T < \Delta T_{th}$, the steady-state current satisfies $J_L > 0$. This is because the forward current driven by bias voltage is greater than the AT induced by temperature gradient, and the net current exhibits conventional transport. However, when $\Delta T > 0$, the AT induced by the temperature gradient will exceed the forward current, so the net current appears as the ANM, and the ANM gradually increases with the increase of the temperature gradient.

In Fig. 5 (b), the temperature gradient $\Delta T = 5$, it can be found that when a forward bias voltage is applied to the system, the system immediately exhibits the ANM. However, when the bias voltage increases, the forward current generated by the bias voltage will gradually offset the AT, and the ANM will disappear when the bias voltage $\Delta V = \Delta V_{th}$. When $\Delta V > \Delta V_{th}$, The steady-state current $J_L$ is always greater than zero and increases with increasing bias voltage.

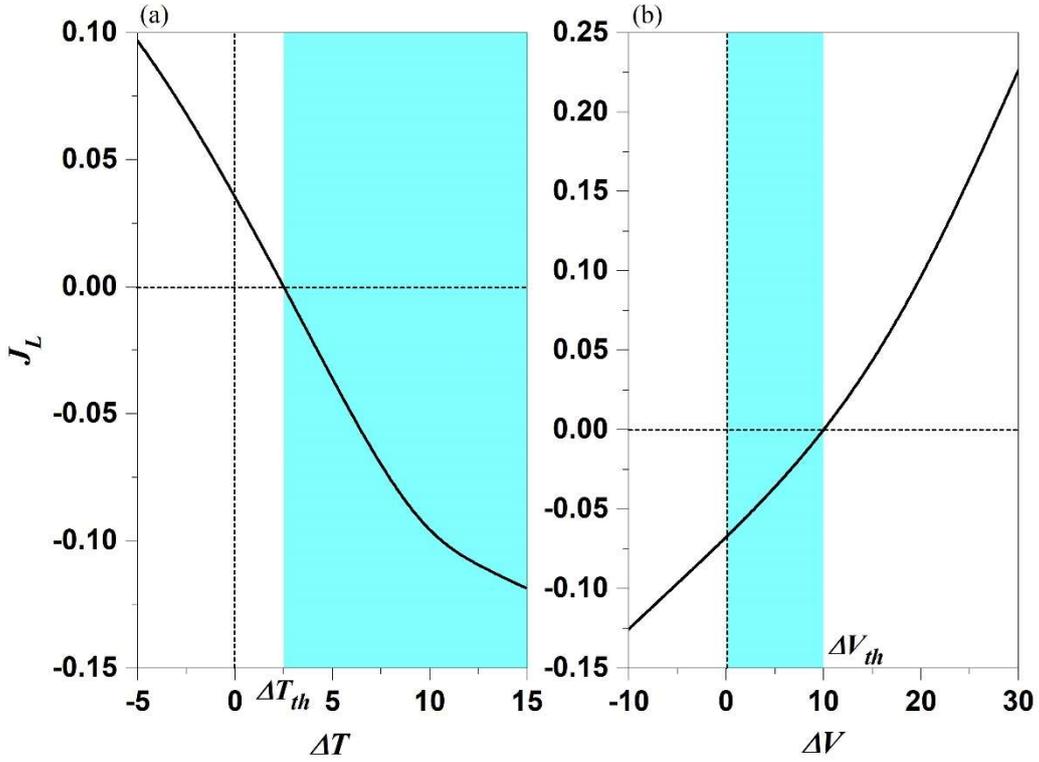

FIG. 5 The steady-state current $J_L$ is depicted as a function of the temperature gradient $\Delta T$ (a) and the bias voltage $\Delta V$ (b), respectively. The absolute negative mobility occurs in blue region. The parameters $T = 7.5$, $\varepsilon = -15$, $\Delta V = 5$ (a), and $\Delta T = 5$ (b).

## V. REVERSE HEAT TRANSPORT

Next, we turn to the steady-state flow $Q_R$ with the temperature gradient $\Delta T = 5$. It can be seen from Fig. 6 that when $\Delta V = 0$, the system is only driven by temperature gradient, and the steady-state heat flow $Q_R$ is always greater than zero, that is, the system heat flow flows from high-temperature reservoir to low-temperature reservoir. However, when the system superimposes a bias voltage ($\Delta V = 2.5$, $\Delta V = 5$) in the same direction as the temperature gradient, in the region $\varepsilon < 0$, the steady-state heat flow $Q_R < 0$ will appear incredibly. This is an anomalous heat transfer, that is, the heat flows from a reservoir with low temperature and low chemical potential to a reservoir with high temperature and high chemical potential. This phenomenon of heat flow is opposite to both applied forces (temperature gradient and bias voltage), which is called as the reverse heat transport.

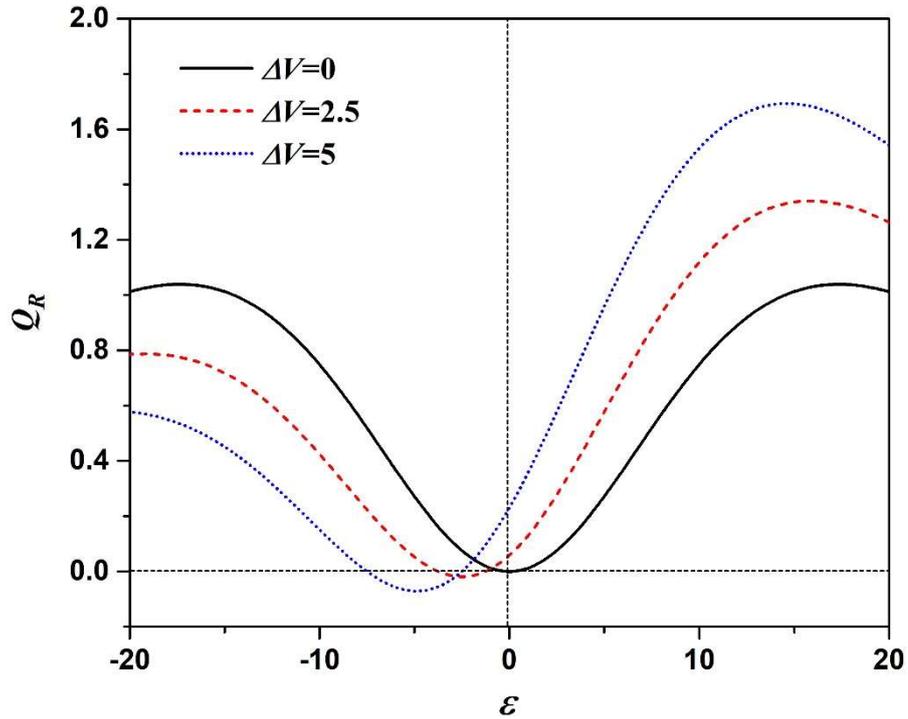

FIG. 6 The steady-state heat flow $Q_R$ is depicted as a function of a single energy level $\varepsilon$ at different bias voltages $\Delta V$. The reverse heat transport occurs in regions where $\varepsilon < 0$ when $\Delta V > 0$. The parameters $T = 7.5$, $\Delta T = 5$.

Fig. 7 shows the curves of the RHT with the bias voltage (a) and temperature gradient (b), respectively. It can be seen from Fig. 7 (a) that when $\Delta T = 5$, the RET occurs in a limited range of bias voltage, and the RHT first increases and then decreases with the increase of bias voltage. Fig. 7 (b) shows that the RHT is induced by the positive bias voltage. When a positive temperature gradient is superimposed on the system, the RHT will be suppressed and eventually the RHT will be transferred to the positive transport with the increase of the temperature gradient.

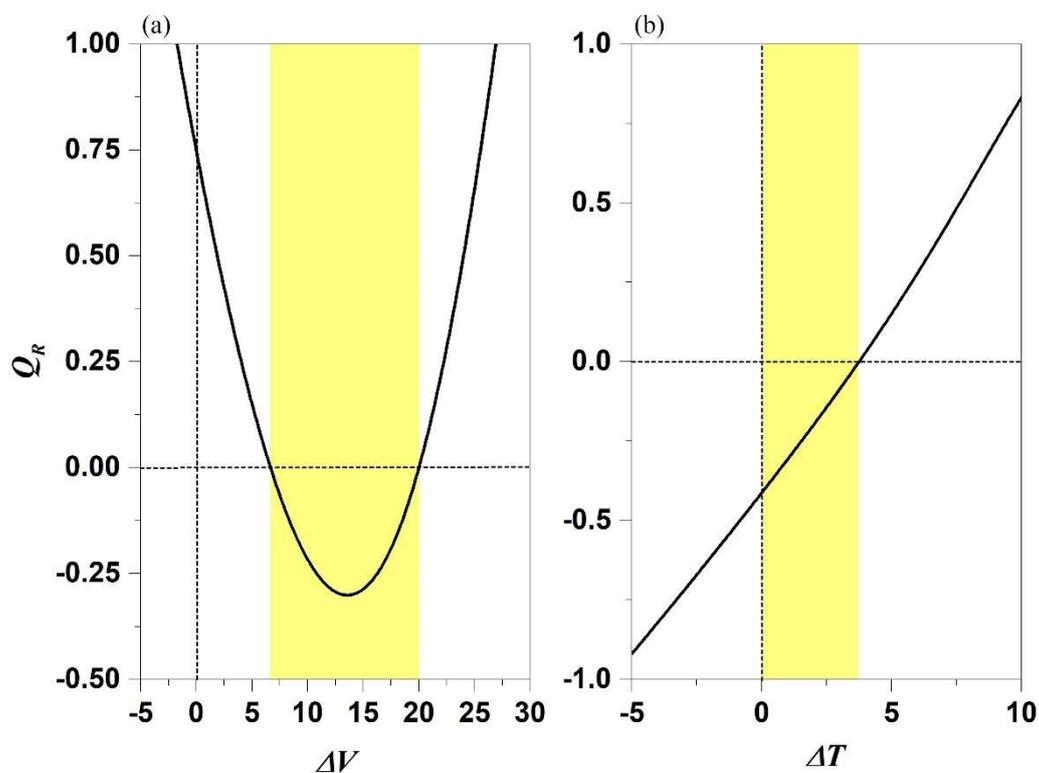

FIG. 7 The steady-state heat flow $Q_R$ is depicted as a function of the bias voltage $\Delta V$ (a) and the temperature gradient $\Delta T$ (b), respectively. The reverse heat transport occurs in yellow region. The parameters $T = 7.5$, $\varepsilon = -10$, $\Delta T = 5$ (a), and $\Delta V = 5$ (b).

Fig. 8 shows the transport characteristics of steady-state heat flow $Q_L$, $Q_R$, steady-state current $J_L$, and steady-state entropy production rate $s$ of the system when $\Delta T = 5$ and $\Delta V = 5$. It can be found that the system has four transport regions with different characteristics, which are labeled with different background colors. In region 1, the steady-state heat flows $Q_L$ and $Q_R$ of system are positive, while the steady-state current $J_L$ of system is negative. This region exhibits ANM. In region 2, $J_L > 0$, while $Q_L < 0$ and $Q_R < 0$. This means that the heat flow in the region flows against the bias voltage and the temperature gradient, that is, the RHT. In region 3, $J_L > 0$, $Q_R > 0$ and $Q_L < 0$. In this region, driven by the bias voltage, work (input power $P = J_L \Delta V$) is converted to heat and transferred to two heat reservoirs. The conventional transport processes are located in region 4. In addition, the entropy production rate $s$ is always greater than zero throughout the region. This means that these counterintuitive phenomena do not violate the second law of thermodynamics. In particular, there is a reversible energy level $\varepsilon_r$ where $s = 0$. Based on Eq. (9), $\varepsilon_r$ is given as

$$\varepsilon_r = \frac{\mu_R T_L - \mu_L T_R}{T_L - T_R}, \tag{10}$$

This reversible energy level distinguishes between the ANM and RHT. When $\varepsilon < \varepsilon_r$, the system shows the ANM, and when $\varepsilon_r < \varepsilon < \mu_R$, the system shows the RHT.

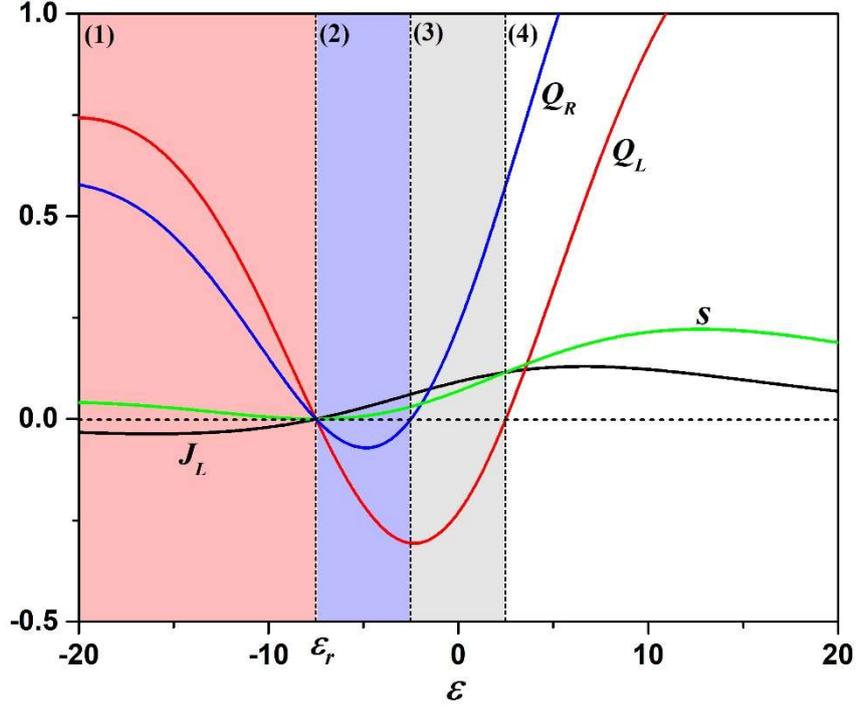

FIG. 8 Anomalous transport. The steady-state heat flow $Q_L$, $Q_R$, the steady-state current $J_L$, and the entropy production rate $s$ are depicted as a function of a single energy level. The parameters $T = 7.5$, $\Delta T = 5$, $\Delta V = 5$.

## VI. DISCUSSIONS

For comparison, Fig. 9 shows the steady-state transport characteristics for $\Delta T = 5$, $\Delta V = -5$. In this case, the transport of the system also exists in four different regions. In region 1, the steady-state heat flow and steady-state current are transported in reverse driven by temperature gradient and bias voltage, respectively. In region 3, work is converted to heat and transferred to two heat reservoirs. In region 3, driven by the bias voltage, the steady-state heat flow from the low temperature heat reservoir to the high temperature heat reservoir, then the system is a thermoelectric refrigerator. In region 4, part of the heat output from the high-temperature heat reservoir is used to drive the current to overcome the bias voltage and do work, and the remaining heat is transported to the low-temperature heat reservoir, which is a thermoelectric heat engine. There is a reversible energy level $\varepsilon_r$ between the thermoelectric heat engine and the refrigerator, which has the same form as Eq. (10). This is a typical thermoelectric transport phenomenon and has been extensively studied [8-12]. Comparing to Fig. 8 and Fig. 9, it can be found that the studies in this paper provide another perspective to understand the thermodynamic properties of steady-state transport of single quantum dot systems.

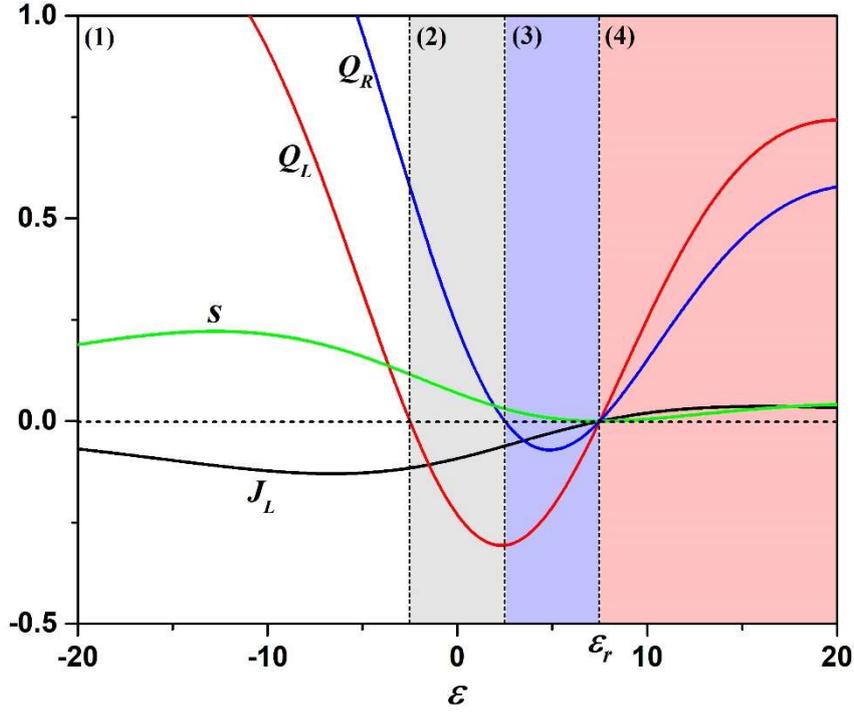

FIG. 9 Thermoelectric transport. The steady-state heat flow $Q_L$, $Q_R$, the steady-state current $J_L$, and the entropy production rate $s$ are depicted as a function of a single energy level. The parameters $T = 7.5$, $\Delta T = 5$, $\Delta V = -5$.

## VII. CONCLUSIONS

We have demonstrated that a single-level quantum dot system deviates from the equilibrium when driven by temperature gradient and bias voltage, it will exhibit counterintuitive transport phenomena, which are AT, ANM and RHT, respectively. These counterintuitive transport phenomena depend on the position of the energy level of single quantum dot, and there is a reversible energy level that distinguishes between absolute negative mobility and reverse heat flow. The results of this paper provide another perspective for understanding the thermodynamic properties of steady-state transport of quantum systems.

**ACKNOWLEDGEMENTS**

This paper is supported by the National Natural Science Foundation of China (Grant No. 12365006 and Grant No. 12304058) and the Natural Science Foundation of Guangxi (China) (Grant No. 2022GXNSFBA035636 and Grant No. 2024GXNSFBA010229)